%% file: main.tex
\documentclass[twocolumn, amsmath, amssymb, aps, prl]{revtex4-1}
\usepackage{graphicx}
\usepackage{dcolumn}
\usepackage{bm}
\usepackage{hyperref}
\usepackage{bibunits}
\usepackage{amsmath}
\usepackage{braket}
\usepackage{mathtools}
\usepackage{array}
\usepackage{float}
\usepackage{epstopdf}
\begin{document}
\title{Proposed Five-Electron Charge Quadrupole Qubit}
\author{John H.~Caporaletti}
\affiliation{Department of Physics, University of Maryland Baltimore County, Baltimore, MD 21250, USA}
\author{J.~P.~Kestner}
\affiliation{Department of Physics, University of Maryland Baltimore County, Baltimore, MD 21250, USA}

\begin{abstract}
A charge qubit couples to environmental electric field fluctuations through its dipole moment, resulting in fast decoherence. We propose the $p$-orbital ($pO$) qubit, formed by the single-electron, $p$-like valence states of a five-electron Si quantum dot, which couples to charge noise through the quadrupole moment. We demonstrate that the $pO$ qubit offers distinct advantages in quality factor, gate speed, readout, and size. We use a phenomenological, dipole two-level-fluctuator charge noise model to estimate a $T_2^* \sim 80$ ns. In conjunction with Rabi frequencies of order $10$ GHz, an order of magnitude improvement in qubit quality factor is expected relative to state-of-the-art semiconductor spin qubits. The $pO$ qubit features all-electrical control via modulating the dot's eccentricity. We also show how to perform two-qubit gates via the $1/r^5$ quadrupole-quadrupole interaction. We find a universal gate set using gradient ascent-based control pulse optimization, subject to $10$ GHz maximum allowable bandwidth and $1$ ns pulse times.
\end{abstract}

\maketitle
\begin{bibunit}[apsrev4-1]
Using qubits to manipulate information fundamentally differs from classical computing schemes and offers potential advantages depending on the computational problem being solved \cite{shor1994,Grover1996,peruzzo_variational_2014}. Many candidate qubit systems exist, such as semiconductor, trapped ion, superconducting, topological, and photonic qubits. Each candidate has unique strengths and weaknesses, populating a high-dimensional parameter space with axes such as size, coherence time, gate speed, scalability, addressability, experimental feasibility, etc. In this work we add a new candidate that prioritizes size and the product of coherence time and gate speed (i.e., quality factor), a quantum dot charge qubit we will refer to as the \emph{$p$-orbital} ($pO$) qubit since it is formed via the $p$-orbital subspace.

This letter shows the $pO$ qubit has several advantages compared to spin qubits. It features baseband all-electrical control via dot deformations and can be operated at zero magnetic field for compatibility with superconducting interconnects. It is compatible with essentially any existing spin qubit device, but avoids the need for micromagnets or microwave magnetic fields. We show below that it allows Rabi frequencies $\sim 10$ GHz, so the gate speed is only limited by the classical control electronics. In addition to reducing the problem size at which quantum advantage emerges, this also has the advantage of reduced noise power at these high frequencies.
Furthermore, it only couples to charge noise through the quadrupole moment, and due to its single-dot construction and lack of leakage states, has a relatively long coherence time which we estimate below to be $T_2^* \sim 80$ ns, potentially allowing an order of magnitude better qubit quality factor metric than any other current semiconductor qubit.

The first realization of a semiconductor charge qubit consisted of a double quantum dot containing a single electron \cite{gorman2005charge,petersson2010quantum}, with logical states being left ($\ket{L}$) or right ($\ket{R}$) dot occupancy. The key feature of this qubit is its large dipole moment between $\ket{L}$ and $\ket{R}$, so it is sometimes referred to as a ``charge dipole" (CD) qubit \cite{Friesen2017}. Accordingly, the CD qubit is strongly coupled to the electric field and therefore has fast gate operations but also fast decoherence. To circumvent decoherence, the CD qubit can be parked at a `sweet spot' corresponding to zero detuning between dot chemical potentials \cite{Kim2015micorwave}, but performing nontrivial gates requires moving the qubit away from the sweet spot. The quality factor, defined as the product of the Rabi frequency and the $T_2^\ast$ dephasing time, for a CD qubit is generally worse than for a spin qubit \cite{Stano2022review}. For this reason, research on semiconductor based qubits in the last 15 years has been dominated by the spin degree of freedom (DOF).

However, a recent proposal, called the charge quadrupole qubit \cite{Friesen2017,kornich2018phonon,koski2020strong,kratochwil2021charge}, moves past the `sweet spot' idea by having no dipole moment between its basis states at \emph{any} operating point. Its coupling to the electric field is always via the quadrupole moment, although this beneficial structuring of the interaction with the environment is at the expense of adding a third dot to the qubit footprint and introducing a low-lying leakage state. Earlier work has also noted the benefit of working with higher-multipole based charge qubits using multiple dots \cite{Oi2005,Harold2007}. The $pO$ qubit, on the other hand, achieves this quadrupole structure with only a single dot and no leakage state. In this sense, the $pO$ qubit marks a distinct evolution of the charge qubit.

In the following, we first present the system Hamiltonian, validate significant assumptions about our model, and discuss single-qubit control. Second, in an effort to quantify the $pO$ qubits performance, we use a dipole two-level fluctuator (TLF) model to estimate the $pO$ qubit's inhomogeneous dephasing time and gate infidelities. We use the same model to estimate corresponding quantities for state-of-the-art semiconductor based qubits in order to draw a fair comparison. Third, we find a universal set of gates via pulse optimization for two neighboring $pO$ qubits with an always-on quadrupole-quadrupole Coulomb interaction. Finally, we show that readout/initialization can be carried out using a quantum point contact to measure along the $x$ or $y$ axes of the Bloch sphere.


Consider a laterally defined quantum dot in a semiconductor heterostructure with an out-of-plane magnetic field of magnitude $B_z$. Modeling the dot as an isotropic 2D harmonic oscillator, the Hamiltonian governing an electron's orbital dynamics is
\begin{equation}\label{Eq: H0}
   H_0 = \frac{(-i\hbar \nabla + e\mathbf{A})^2}{2m^*} + \frac{1}{2}m^* \omega_0^2 r^2,
\end{equation}
where the symmetric gauge is chosen, $\mathbf{A}=\frac{B_z}{2}(-y\hat{\textbf{x}}+x\hat{\textbf{y}})$, yielding $\textbf{B} = B_z \hat{\textbf{z}}$. Additionally, $r$ is the radial position operator, $\omega_0$ is the confinement frequency, $m^*$ is the effective mass and $e$ is the electron's charge. Solving the corresponding Schr\"{o}dinger equation with the Hamiltonian in Eq.~\eqref{Eq: H0} yields the Fock-Darwin states \cite{Fock1928},\cite{Darwin1927}
\begin{equation}\label{Eq: state}
    \begin{aligned}
        \psi_{n,m}(x,y)=&\frac{1}{l_0}\sqrt{\frac{\left(\frac{n-|m|}{2}\right)!}{\pi\left(\frac{n+|m|}{2}\right)!}} \left(\frac{x + i y \text{sgn} (m)}{l_0}\right)^{|m|} \\
        & \times e^{-\frac{x^2 + y^2}{2 l^2_0}} L^{|m|}_{\frac{n-|m|}{2}}\left(\frac{x^2 + y^2}{2 l^2_0}\right),
    \end{aligned}
\end{equation}
where $l_0=l_B/(\frac{1}{4}+\frac{\omega_0^2}{\omega_c^2})^\frac{1}{4}$, $l_B = \sqrt{\hbar/m^*\omega_c}$, and $\omega_c = eB_z/m^*$ are the dot's characteristic length, magnetic length and cyclotron angular frequency respectively. $L^a_b(x)$ is the associated Laguerre polynomials, $n \in\mathbb{Z}_0^+$, and $m = -n, -n+2, \ldots, n-2, n$. The energy spectrum is
\begin{equation}\label{Eq: Energy Spectra}
    E_{n,m}= (n+1)\hbar \omega_c \sqrt{\frac{1}{4}+\frac{\omega_0^2}{\omega_c^2}} +m\frac{\hbar}{2} \omega_c.
\end{equation}
In analogy with the hydrogen atom, $n=0$ is referred to as the $s$ orbital, $n=1$ as the $p$ orbitals and so on. Specifically, $\psi_{1,m}(x,y) =\braket{x,y|p_m}$ and $\{\ket{p_{+1}}, \ket{p_{-1}}\}$ spans a two-dimensional subspace that is energetically isolated from the $s$ and $d$ orbital states by the orbital level spacing, given that $\omega_0\gg\omega_c$.

Realistically, in a Si device, one must also take the valley and spin DOF into account, so the $p$-orbital subspace contains eight levels. However, the Rashba and Dresselhaus-like spin-orbit-coupling (SOC) \cite{Bihlmayer_2015,dresselhaus_spin-orbit_1955} Hamiltonians within the $p$-orbital subspace are exactly zero because the matrix elements have integrands with odd parity, preventing leakage from the orbital encoding to spin. Valley-orbit coupling (VOC) can exist in the presence of devices that have interface steps or Ge alloy disorder in the well \cite{friesen_theory_2010, gamble_disorder-induced_2013, losert_practical_2023, mcjunkin_valley_2021}. For simplicity, we assume negligible VOC for the remainder of the paper. However, if VOC is non-negligible, the fundamental idea does not change, as shown in the Supplemental Material (SM) \cite{sup1}.

In order to have an electron stably occupy the $p$ orbital, one must fill the $s$ orbital such that orbital decays are Pauli blocked. Accounting for the spin and valley DOF, one needs four electrons to fill the four $s$ spin-orbital-valley states, and a fifth electron to reside in the $p$ orbital. This idea of a `frozen core' \cite{Sabelli1975,Hu2001,Barnes2011} approximates the intra-dot e-e interactions as a renormalization to the dot potential size, generally preserving the $p$ orbital's characteristic shape and the energetic separation between the $p$-orbital manifold and other orbital manifolds, and the intuitive idea has both theoretical \cite{liang_electronic_2024} and experimental support \cite{Leon2020}.

\begin{figure}[ht]
\centerline{\includegraphics[width=\columnwidth]{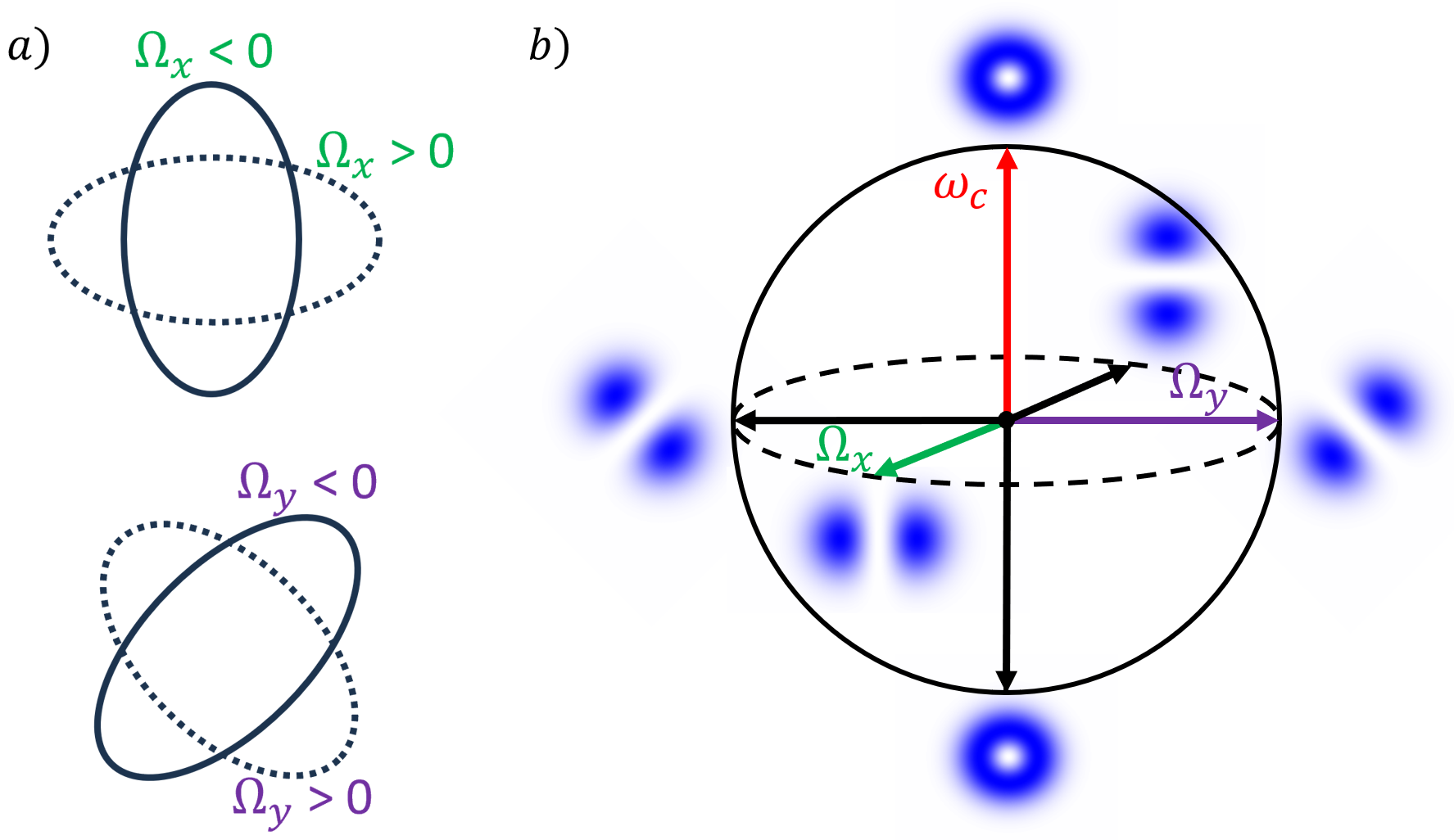}}
\caption{a) Various potential deformations and the resulting types of rotations. b) Qubit rotation axes, with $pO$ charge distribution shown at six of the Bloch sphere's antipodes.}
\label{Fig: Bloch Sphere}
\end{figure}

Qubit operations within this subspace can be performed via deformations of the dot, similar to the case of a spin-charge qubit \cite{kyriakidis2007universal}. Consider a deforming potential operator, $V_d$, expanded about the center of the dot, 
\begin{equation}\label{Eq: V def}
    V_{d} \approx V_d|_{0} +\left(\partial_i V_d\right)|_{0} r_i + \frac{1}{2} \left(\partial_j\partial_k V_{d}\right)|_{0} r_j r_k,
\end{equation}
where $r_a \in \{x,y\}$, $\partial_a \in \{\partial_x,\partial_y\}$ and repeated indices are summed over. Writing the corresponding deformation Hamiltonian in the $\ket{p_m}$ basis yields
\begin{equation}\label{eq: V def p basis}
    \begin{aligned}
        H_d = eV_d &\approx \frac{el_0^2}{4}\left((\partial_x^2-\partial_y^2)V_d|_{0} \sigma_x+2\partial_x\partial_y V_d|_{0} \sigma_y\right)\\ &=\frac{\hbar}{2} \left(\Omega_x \sigma_x + \Omega_y \sigma_y\right),
    \end{aligned}                                  
\end{equation}
where $\sigma_i$ are the Pauli operators in this basis. The monopole and dipole term of Eq.~\eqref{Eq: V def} contribute identity and zero respectively. Therefore, the deforming Hamiltonian is dictated, to leading order, by the quadrupole terms shown in Eq.~\eqref{eq: V def p basis}. Each term can be interpreted as a difference in confinement frequencies along a pair of orthogonal spatial axes of the dot (the two sets differing by an angle of $\pi/4$) as shown in Fig.~\ref{Fig: Bloch Sphere}a. Controllable dot anisotropies have been demonstrated corresponding to a difference in confinement energies along orthogonal axes of order $100 \mu\text{eV}$ \cite{Leon2020}, implying that $\Omega_i$ can range from zero (for an isotropic dot) up to $\sim 2\pi \times 25$ GHz. For a dot size of $l_0 = 10$ nm (which we take henceforth), this corresponds to a deformation of $\sim 1\%$. Even if tuning $\Omega_i$ to zero (i.e., an isotropic dot) is infeasible experimentally, any amount of tunability permits independent axes of rotation and thus full single-qubit control (most efficiently if the range of control over the anisotropy is not much smaller than the inherent anisotropy).

The single-qubit $pO$ Hamiltonian in the $\ket{p_m}$ basis is thus
\begin{equation}\label{Eq: Hsub}
    H_{1q}=H_0+H_d=\frac{\hbar}{2}\left(\Omega_x \sigma_x + \Omega_y \sigma_y +\omega_c \sigma_z\right),
\end{equation}
with $\Omega_x,\Omega_y, \omega_c \in \mathbb{R}$. A graphical depiction of Eq.~\eqref{Eq: Hsub} on the Bloch sphere is shown in Fig.~\ref{Fig: Bloch Sphere}b. Only two of the three axes in Eq.~\eqref{Eq: Hsub} are required to span $\mathfrak{su}(2)$. If $\omega_c \neq 0$, one type of deformation suffices. If we turn off the magnetic field entirely (i.e., $\omega_c = 0$), independent control over $\Omega_x$ and $\Omega_y$ is required. A gate electrode architecture such as the crossbar design developed in Ref.~\cite{li2018crossbar,Borsoi2024} is sufficient for this. Working with zero magnetic field is desirable for compatibility with superconducting resonator interconnects, but an always-on magnetic field can serve to partially continuously decouple charge noise, similar to Ref.~\cite{nichol_high-fidelity_2017}. We show below that the decoherence time at zero magnetic field is already relatively high, so one may safely choose to work at zero field if desired.


The control terms $\Omega_i$ are a result of the $pO$ qubit's coupling to the electric field, but this also facilitates decoherence. However, the coupling is now via the quadrupole moment to second derivatives of the electric potential instead of via the dipole moment to first derivatives of the electric potential. We expect that as the order to which the qubit couples increases, our ability to artificially induce these structured changes in the environment (e.g., deforming the dots instead of simply detuning them relative to each other) will increasingly exceed the natural level of fluctuations at that order. To demonstrate this, and compare our qubits decoherence to the state-of-the-art, we now use a phenomenological model of charge noise.

Inhomogeneous dephasing will be estimated by $T_2^* = \frac{\sqrt{2}\hbar}{\sigma_E}$ \cite{kubo1954note,petersson2010quantum,Kawakami2014Electrical},
where $\sigma_E = \sqrt{\braket{E^2}-\braket{E}^2}$ with $E$ being the energy splitting between eigenstates of the system's full Hamiltonian, including noise, and $\braket{\cdot}$ denoting an ensemble average. Additionally, relaxation can occur due to phonons. Although typical orbital relaxation times are rather fast, $T_1 \approx 100$ ps \cite{TahanRelaxation2014}, relaxation within the $p$-orbital subspace is estimated in the SM \cite{sup1} to be much slower, $T_1 \approx 1$ s. Conceptually, this originates from the relatively smaller energy splitting of the $p$ orbitals. This promotes coupling to longer wavelength phonons, which in turn creates weaker deformation of the lattice within the quantum dot and therefore slower relaxation.

Charge noise is suspected to originate from an ensemble of TLFs \cite{Kuhlmann2013Charge}.  Recent work suggests that the TLFs consist of electric dipoles at, and oriented parallel to, the interface between the heterostructure and the electrodes \cite{Connors2019}. This ensemble of dipoles produces a potential $V_{\text{noise}}(x,y)$ at the dot, producing an additional uncontrolled source of deformation in Eq.~\eqref{eq: V def p basis}. The energy splitting $E$ is
\begin{equation}\label{Eq: Noise Energy}
\begin{aligned}
    E &=||H_{1q}+eV_{\text{noise}}||
    \\
    &=\hbar \sqrt{(\Omega_x+\delta \Omega_x)^2+(\Omega_y+\delta \Omega_y)^2+\omega_c^2},
\end{aligned}
\end{equation}
where $||A|| = \sqrt{Tr(A A^\dagger)}$.
Shown in the SM \cite{sup1}, we perform several Monte Carlo simulations with the parameters given in Table \ref{Table: noise params} to obtain an estimate of $T_2^* \approx 80$ ns. All simulations use $\omega_c=0$, as this is when the $pO$ qubit is most vulnerable to charge noise while idling.
\begin{table}[ht]
\centering
\begin{tabular*}{\columnwidth}{l c c}
 \hline
 Parameter & Symbol & Value \\
 \hline
  Dipole Length& $d$ & $.1$ nm \cite{reinisch2006microscopic} \\ 
  Dot Distance to Electrodes& $D$ & $100$ nm \\ 
  Characteristic Dot Length& $l_0$ & $10$ nm \\ 
  Dipole Density& $\rho$ & $10^{-3} \text{nm}^{-2}$ \cite{Zimmerman1981}\\
  Interface Area& $A$ & $10^4 \text{nm}^2$\\
  \hline
\end{tabular*}
\caption{Noise model parameters.}
\label{Table: noise params}
\end{table}

The above only accounts for charge noise coupling to the quadrupole moment. There can also be dephasing due to magnetic Overhauser noise and higher order electric coupling. The magnetic coupling of the $pO$ qubit differs from a typical Loss-DiVincenzo (LD) style qubit \cite{LD1997} because the orbital gyromagnetic ratio features the effective electron mass in Si while the spin gyromagnetic ratio uses the intrinsic electron mass. Consequently, the magnetic dephasing of a $pO$ qubit is proportional to that of an LD qubit, $T_2^*=\frac{g m^*}{2 m_e} T_2^{*(LD)}$, where $g$ is the electron $g$ factor and $m_e$ is the bare electron mass. Using the inhomogeneous dephasing time of an LD qubit in isotopically enriched silicon $T_2^{*(LD)}=33 \mu$s \cite{Dzurak2018} and $m^* = 0.19 m_e$ \cite{burkard2023}, the $pO$ magnetic dephasing time is $ \approx 6 \mu \text{s}$. Therefore, Overhauser noise is not the limiting factor in dephasing and can safely be neglected.

As for higher order electric noise, the $pO$ qubit does not have any coupling to odd spatial derivatives of the potential about the dot because the $p$ orbitals are symmetric about the origin. Therefore, to find corrections to our model, one must look past the octupole moment to the hexadecapole moment, which couples to the $4^{th}$ order derivatives of $V_{\text{noise}}$. Assuming the potential produced by a single point dipole, the ratio between the quadrupole and $n^{th}$ term in the Taylor expansion of Eq.~\eqref{Eq: V def} goes like $\sim (l_0/D)^{n-2}$ after averaging over all possible dipole locations and orientations in $A$. The hexadecapole to quadrupole ratio is then $\sim 10^{-2}$ for the parameters in Table~\ref{Table: noise params}. Therefore, higher order electric couplings can be safely ignored.


One of the simplest ways to quantify a qubit's performance is the quality factor, given by
\begin{equation}\label{Eq: Q}
    Q=f_R T_2^*,
\end{equation}
where $f_R$ is the Rabi frequency accessible to the qubit. Eq.~\eqref{Eq: Q} is in accordance with the \emph{qubit} quality factor in Ref.~\cite{Stano2022review} and represents an intrinsic qubit quality, i.e., in the absence of any dynamical decoupling. This is in contrast to the \emph{gate} quality factor, which measures dephasing when driving on an axis orthogonal to the quantization axis. Assuming the validity of the TLF ensemble noise model, we predict the $pO$ qubit to have $Q \approx 1000$.
The highest semiconductor single-qubit quality factor we know of is for an LD spin qubit controlled via electric dipole spin resonance (EDSR) with $Q=78$ \cite{Yoneda2018}. Applying our noise model to the LD spin qubit using the parameters of Ref.~\cite{Yoneda2018}, we obtain the experimentally observed value of $T_2^* \approx 20 \text{ }\mu s$ \cite{sup1} with a confinement energy $\hbar \omega_0 \approx 2$ meV, in agreement with the order of magnitude estimation in Ref.~\cite{Yoneda2018}. Using their estimate $\hbar \omega_0 = 1$ meV directly without any fitting predicts a shorter dephasing time of $\sim 5 \text{ }\mu$s, so our estimate for the $pO$ qubit of $Q \simeq 1000$ may be overly conservative.


We also need multi-qubit interactions for entangling operations. Consider two adjacent quantum dots whose centers lie at $x=\pm \frac{L}{2}$. In the limit that $L \gg l_0$, the single-particle basis remains harmonic oscillator states,
\begin{equation}\label{Eq: Shifted Single Particle State}
    \psi_{1,m}^{\pm}(x,y)=\braket{x,y|p_m^\pm}=e^{\pm i \frac{L y}{4l_b^2}}\psi_{1,m}(x\pm \frac{L}{2},y),
\end{equation}
where the $+ (-)$ sign corresponds to occupancy of the left (right) dot. The phase factor is the result of a gauge transformation when redefining the coordinate system's origin to be mid-way between dots, and vanishes in the absence of a magnetic field \cite{Barnes2011}.

Although the complete two-electron state must be antisymmetric, we will consider distances of $L$ such that wavefunction overlap produces negligible kinetic exchange compared to the Coulomb interaction. Kinetic exchange is undesirable because its strength depends on both the joint spin state and orbital state of both electrons, thus coupling the spin and orbit DOFs and ultimately producing decoherence in the $p$ orbital space. When such wavefunction overlap is negligible, we can treat electrons in each dot as distinguishable. Therefore, the two particle charge basis states subsequently used are 
\begin{equation}\label{Eq: 2 particle state}
    \ket{m,m'}=\ket{p_m^{+}}_1 \otimes \ket{p_{m'}^{-}}_2,
\end{equation}
where the subscripts $1$ and $2$ distinguish the electrons.

\begin{figure}[ht]
    \centering
    \includegraphics[width=\columnwidth]{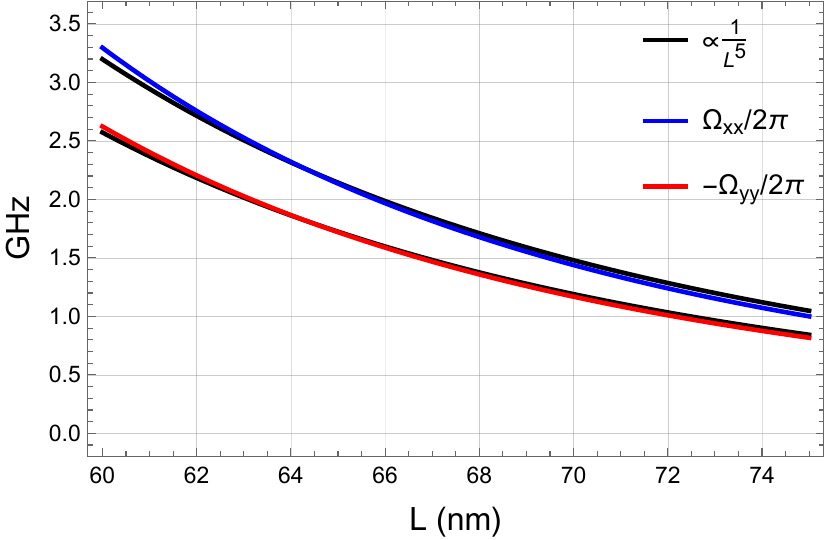}
    \caption{Two-qubit interaction strengths vs inter-dot distance $L$. $\frac{1}{L^5}$ fit is shown in black for both $\Omega_{xx}$ and $\Omega_{yy}$.}
    \label{Fig: fxx contour plot}
\end{figure}

The two-particle Coulomb interaction Hamiltonian, written in the basis states of Eq.~\eqref{Eq: 2 particle state} for two pO qubits a distance $L$ apart is
\begin{equation}\label{Eq: Hc charge basis}
    H_c=\frac{\hbar}{2} \left(\Omega_{xx} \sigma_x^{(1)} \sigma_x^{(2)} +\Omega_{yy} \sigma_y^{(1)} \sigma_y^{(2)} + a \left(\sigma_x^{(1)} + \sigma_x^{(2)}\right)\right),
\end{equation}
where $\Omega_{xx}$ and $\Omega_{yy}$ are the two-qubit quadrupole-quadrupole Coulomb interaction strengths, $a$ is the single-qubit effect of deformation of one dot due to the charge monopole of the other, and the superscripts denote which part of the tensor space the Pauli operators act on. $\Omega_{xx}$ and $\Omega_{yy}$ are opposite in sign and shown in Fig.~\ref{Fig: fxx contour plot}.
For the domain of Fig.~\ref{Fig: fxx contour plot}, $a$ ranges from approximately $5-20$ GHz, effectively skewing the origin of $\Omega_x^{(1),(2)}$. However, this effect would be calibrated away in the initial setup of the device. No $\sigma_z$ terms are present in Eq.~\eqref{Eq: Hc charge basis} because the Coulomb operator cannot distinguish between the identical charge distributions of $\ket{p_\pm}$. Additionally, the term $\sigma_x \sigma_y$ is not present because the $\sigma_x$ eigenstates, $\ket{p_{x,y}}$, do not produce a difference in curvature along the two axes $\hat{x} + \hat{y}$ and $\hat{x} - \hat{y}$. The full Hamiltonian, combining both local and interaction energies, becomes
\begin{multline}\label{Eq: H}
        H_{2q} = \sum_{k=1}^2 \frac{\hbar}{2}\left(\Omega_x^{(k)} \sigma_x^{(k)} + \Omega_y^{(k)} \sigma_y^{(k)} + \omega_c \sigma_z^{(k)}\right) 
        \\
        + \frac{\hbar}{2}\Omega_{xx} \sigma_x^{(1)} \sigma_x^{(2)} +\frac{\hbar}{2}\Omega_{yy} \sigma_y^{(1)} \sigma_y^{(2)}.
\end{multline}


The always-on Coulomb interaction in Eq.~\eqref{Eq: H} complicates single-qubit gates. One option is to use shuttling to change the distance between qubits and turn the interaction on/off, or use VOC as mentioned in the SM \cite{sup1}, but even with always-on interaction it is possible to use shaped pulses to obtain universal control. Using gradient ascent pulse engineering (GRAPE) \cite{Khaneja2005Optimal}, we search the space of possible pulses for qubit unitaries within the universal set $\{H^{(k)}, S^{(k)}, T^{(k)}, \text{bSWAP}\}$, where $H^{(k)}$ is the Hadamard gate on qubit $k$, and likewise $S^{(k)}$ and $T^{(k)}$ are phase gates of angle $\frac{\pi}{2}$ and $\frac{\pi}{4}$ respectively, while $\text{bSWAP} = e^{\left(i~\pi/4~(\sigma_x^{(1)}\sigma_x^{(2)}-\sigma_y^{(1)}\sigma_y^{(2)})\right)}$ \cite{wei2024native} is a perfectly entangling gate \cite{zhang2003geometric} and is equivalent to $\text{iSWAP} = e^{\left(i~\pi/4~(\sigma_x^{(1)}\sigma_x^{(2)}+\sigma_y^{(1)}\sigma_y^{(2)})\right)}$ up to a local $\sigma_x$ rotation \cite{poletto2012entanglement}. The pulse control parameters are given by the set $C = \{\Omega_x^{(1)},\Omega_y^{(1)},\Omega_x^{(2)},\Omega_y^{(2)}\}$. Gate times are $1$ ns for all pulses \cite{sup1} with system parameters $L=63$ nm, corresponding to $\Omega_{xx} \approx 2.5$ GHz and $\Omega_{yy} \approx -2.0$ GHz. Our choice of $L$ is large enough that $\Omega_{xx,yy}$ is more than $100 \times$ greater than kinetic exchange, as assumed in Eq.~\eqref{Eq: 2 particle state}, but the two dots are still near enough that $\Omega_{xx,yy} \sim \Omega_{x,y}$ so two-qubit gates can be performed on the same timescale as single-qubit gates. The approximate range of $L$ which satisfies these requirements is $60 \lesssim L\lesssim 75$ nm. The pulses assume an available bandwidth of at most 10 GHz, which is the state of the art for arbitrary waveform generators (AWG) \cite{Kim2014Quantum}. Similar optimizations could also be performed for less advanced AWGs as necessary.

For each gate in the universal gate set, we again perform Monte Carlo noise simulations using the parameters in Table \ref{Table: noise params}, taking into account both single-qubit noise (i.e., fluctuations in eccentricity, causing noise in $\Omega_x$ and $\Omega_y$) and two-qubit noise (i.e., fluctuations in the inter-dot distance $L$, causing noise in $\Omega_{xx}$ and $\Omega_{yy}$). However, for $L=63$ nm and $l_0 = 10$ nm, single-qubit noise is the limiting factor. Infidelities are found to be $\sim 10^{-5}-10^{-4}$ \cite{sup1}.

\begin{figure}[ht]
    \centering
    \includegraphics[width=\columnwidth]{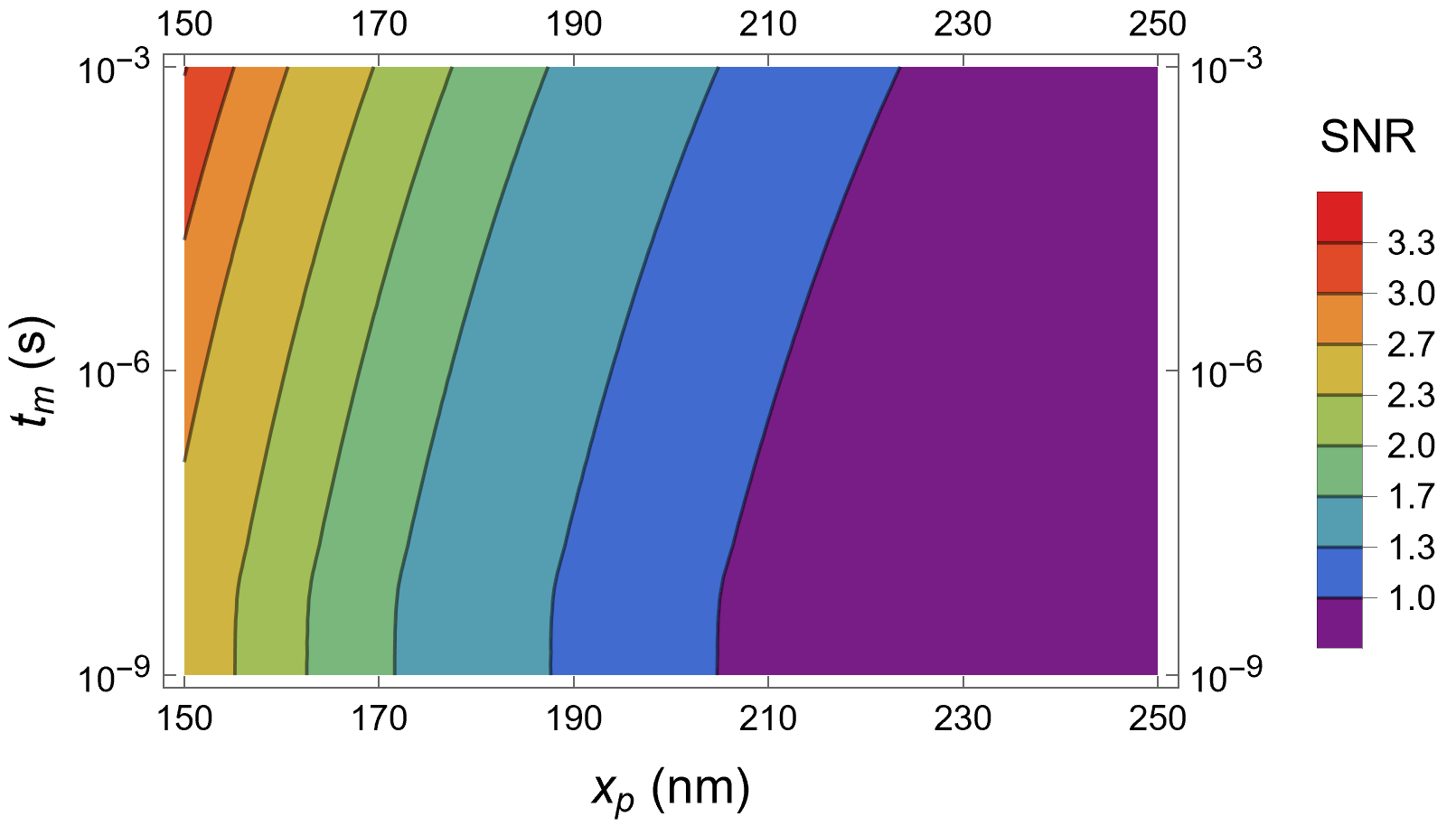}
    \caption{Signal-to-noise ratio as a function of QPC-to-qubit distance ($x_p$) and measurement time ($t_m$).}
    \label{Fig: SNR}
\end{figure}

We now consider initialization and readout for the $pO$ qubit. Recognizing that the eigenstates of $\sigma_x$, $\ket{p_{x,y}}$, have different charge configurations (see Fig.~\ref{Fig: Bloch Sphere}a), it is possible to perform readout on the qubit state without the selective tunneling typically used in Elzerman readout \cite{Elzerman2004} or Pauli spin blockade \cite{lai2011pauli}. Different voltages are produced by $\ket{p_{x}}$ and $\ket{p_{y}}$ at a device such as a quantum point contact (QPC) nearby. This corresponds to a chemical potential difference at the QPC and can be extracted using current measurements, as in Refs.~\cite{Connors2019,PaqueletWuetz2023}. For charge noise with a power spectral density $S=\frac{A}{f}$ and $A\sim 1\,\mu\text{eV}^2$ \cite{Connors2019,connors2022,Freeman2016Comparison,Struck2020Low,PaqueletWuetz2023},
we calculate the signal-to-noise ratio (SNR) as a function of QPC-to-qubit distance $(x_p)$ and measurement time $(t_m)$ using $SNR=\frac{\braket{V}}{\sigma_V}$, where $V=\int_{0}^{t_m}dt V(t)$ and $V(t)$ is the measured voltage at the QPC as a function of time. Fig.~\ref{Fig: SNR} shows that the $pO$ qubit offers rapid readout.


In summary, we propose a charge qubit that uses the two-level $p$ orbital subspace in a five-electron silicon quantum dot. We found an all-electrical, universal gate set with $1$ ns gate times, with similarly fast readout, and even faster gate speeds $\sim 100$ ps are feasible if one has fast enough classical control electronics. We estimate an inhomogeneous dephasing time of the $pO$ qubit of $T_2^* \approx 80$ ns, implying a single-qubit quality factor of $Q\sim1000$. Looking forward, the $pO$ qubit appears to be a strong direction for future research because it requires only a single dot, with no low-lying leakage states, and electric control. Future work will investigate the use of finite VOC to enhance coherence times and reduce device complexity, along with developing a scalable array architecture of $pO$ qubits. 

\begin{acknowledgements}
    \textit{Acknowledgments---} The authors acknowledge support from the Army Research Office (ARO) under Grant Number W911NF-23-1-0115. Additionally, the authors thank Dr. David Kanaar for helpful discussions during the early stages of this work.
\end{acknowledgements}

\begin{acknowledgements}
    \textit{Data Availability---} The data that support the findings of this Letter are openly available \cite{GitHub}
\end{acknowledgements}

\putbib[ref_clean]
\end{bibunit}

\input{sup_v2}

\end{document}

%% file: sup_v2.tex
\onecolumngrid
\clearpage
\begin{center}
\textbf{\large Supplemental Material}
\end{center}

\renewcommand{\theequation}{S\arabic{equation}}
\renewcommand{\thefigure}{S\arabic{figure}}
\renewcommand{\bibnumfmt}[1]{[S#1]}
\renewcommand{\citenumfont}[1]{S#1}
\begin{bibunit}[apsrev4-1]
\section{I. \ Disordered Device and Valley-Orbit-Coupling}\label{VOC}
 The $p$-orbital states can have different valley splittings in realistic devices because the charge distributions are largely non-overlapping and therefore sample different heterostructure features such as alloy disorder or interface steps, leading to valley-orbit coupling (VOC). In the main text, we neglect VOC so the eigenstates of the valley-orbit system are separable. With finite VOC, the eigenstates become hybridized, but we can still use them to encode the qubit as before. In general, we expect a VOC Hamiltonian of the form
\begin{equation}\label{eq: VOC}
    H_{\text{VOC}}=A_{xx}\sigma_x\tau_x + A_{yy}\sigma_y\tau_y + A_{xy}\sigma_x\tau_y + A_{yx}\sigma_y \tau_x,
\end{equation}
where each generator coefficient is an independent random variable (over the space of possible heterostructures), $\sigma$ belongs to the orbital DOF, $\tau$ belongs to the valley DOF, and no $\sigma_z$ terms are present because its eigenstates, $\ket{p_\pm}$, have identical charge distributions.

\begin{figure}[h]
    \centerline{\includegraphics[width=.7\columnwidth]{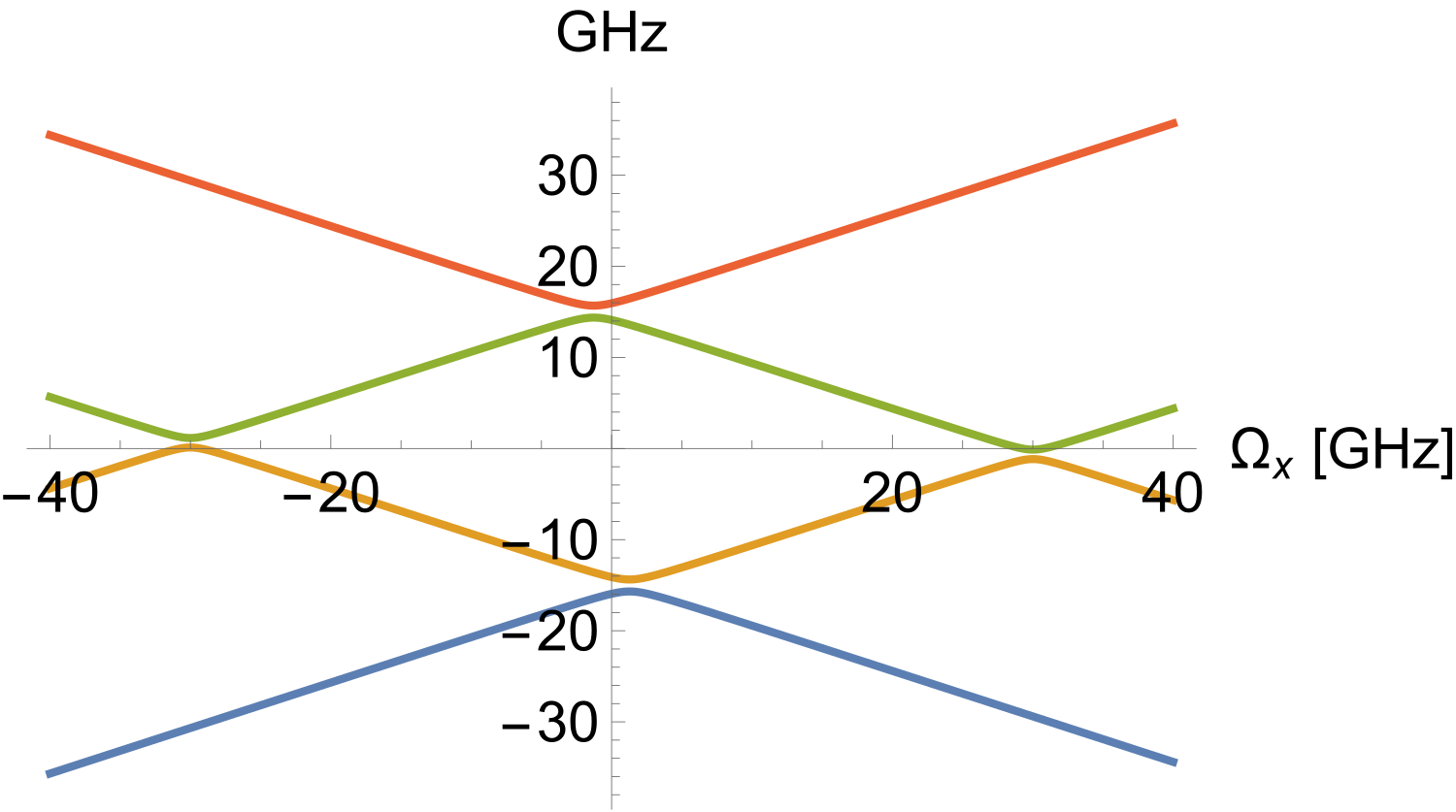}}
    \caption{Instantaneous eigenenergies of the valley-p-orbital manifold for $\Omega_y = 0$, $\{A_{xx},A_{yy},A_{xy},A_{yx}\} = \{0.639, 0.521, 0.778, -0.658\} \text{ GHz}$. The orbital-independent valley splitting is chosen to be $2\Delta_{vs} \sim 150 \text{ }\mu\text{eV} \sim 30$ GHz \cite{losert_practical_2023}.}
    \label{fig:eigen}
\end{figure}

To estimate the generator coefficients in Eq.~\eqref{eq: VOC} we use the discretized, effective mass model of Ref.~\cite{losert_practical_2023} and consider only alloy disorder because it dominates the valley splitting physics for heterostructure interfaces of realistic width. As in Ref.~\cite{losert_practical_2023}, the Si/Ge concentration at each discretized cell is modeled as a binomial random variable whose mean value follows a sigmoid function about each interface. Considering a discretized cell size of $a_0/8$, containing one atom, and a realistic interface width of $\lambda_{\text{int}} = 1$ nm \cite{degli_esposti_low_2024}, every generator coefficient in Eq.~\eqref{eq: VOC} has a mean of zero and a standard deviation of $\sigma \sim 1$ GHz. In Fig.~\ref{fig:eigen}, the four valley-$p$-orbital eigenenergies are plotted as a function of the deformation-induced $p$-orbital splitting, $\Omega_x$, and $\{A_{xx},A_{yy},A_{xy},A_{yx}\} = \{0.639, 0.521, 0.778, -0.658\}$ which are sampled points from a Gaussian with zero mean and standard deviation $\sigma$.

In the presence of finite VOC, the degeneracy between the computational states considered in the main text, $\ket{p_{\pm}}\otimes\ket{v_-}$, where $\ket{v_-}$ is the low energy eigenstate of orbital-independent valley Hamiltonian, is lifted at $\Omega_{x,y}=0$. This adds a static control term along the equator of the Bloch sphere. One also still has two linearly independent axes of control in the two types of deformations, $\Omega_{x,y}$. In this encoding, the region $\Omega_{x,y} \sim \Delta_{vs}$ must be avoided to prevent leakage into the excited valley manifold via the outer avoided crossings.

Not only does such a hybridized $pO$ qubit continue to work in the presence of VOC just as before, but it actually carries additional benefits. Foremost, given the same encoding, operations could be performed in a Landau-Zener style by manipulating only $\Omega_x$, diabatically ramping through the avoided crossing to perform a $Z$ rotations and parking at non-zero $\Omega_x$ to perform $X$ rotations, akin to Ref.~\cite{koh_pulse-gated_2012}. This lightens the necessary gate architecture complexity for implementing the $pO$ qubit in a real device, while still operating with zero magnetic field. Additionally, this avoided crossing opens up the possibility of a quadrupolar charge-noise sweet spot (akin to Ref.~\cite{kim_microwave-driven_2015}), leading to further improved decoherence properties. Finally, if one chooses to encode the qubit in the $\{\ket{p_{+}}\otimes\ket{v_-}, \ket{p_{-}}\otimes\ket{v_+}\}$ subspace (spanned by the middle two eigenstates in Fig.~\ref{fig:eigen}), two-qubit operations are effectively turned off when parked at one of the outer avoided crossings due to an effective ``smearing out" of the $p$-orbitals from their hybridization with the valleys. These three aspects indicate that the presence of non-negligible device disorder, while requiring a careful characterization of each dot's spectral response to deformation, would provide a path towards further improvement of the $pO$ qubit.

\section{II. \ Monte Carlo}\label{Monte}
We estimate $T_2^*$ via a Monte Carlo noise simulation. We model the noise source as $n = \rho A$ dipoles at the heterostructure interface and oriented parallel to it. First, the center of mass position $(x_i,y_i)$ and in-plane orientation $(\phi_i)$ of the dipoles are randomly and uniformly generated over the intervals $\{x_i,y_i \in \mathbb{R} | -\frac{\sqrt{A}}{2}<x_i,y_i<\frac{\sqrt{A}}{2}\}$ and $\{\phi_i \in \mathbb{R} | 0<\phi_i<2\pi\}$. This randomly selects a single configuration of dipoles out of the ensemble of possible configurations, defining a particular, random noise potential instance. Then we calculate $\delta \Omega_x$ and $\delta \Omega_y$ by taking second derivatives of the potential instance at the dot to calculate the dot deformation, we calculate the fluctuated eigenenergy splitting in the p-orbital subspace. Repeating this process $N$ times, we generate a distribution of energy splittings, from which we obtain a standard deviation $\sigma_E$ and hence, $T_2^*$.

\begin{figure}[h]
    \centerline{\includegraphics[scale=.5]{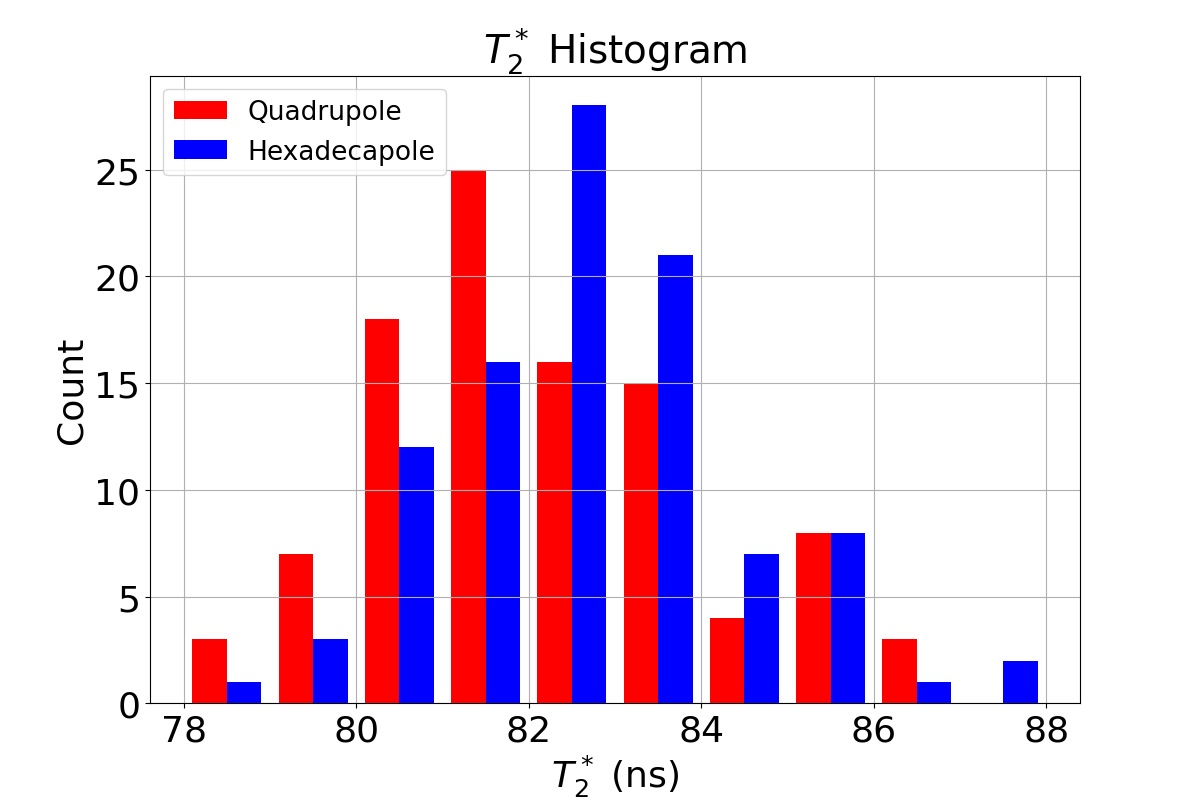}}
    \caption{Two $T_2^*$ distributions plotted simultaneously. Red data corresponds to the noise potential truncated at the quadrupole order, while the blue is truncated at the hexadecapole order.}
    \label{Fig: T2 dist}
\end{figure}

$T_2^*$ times reported in the main text represent a mean $T_2^*$ time for 100 Monte Carlo simulations with $N=1000$, as indicated in Fig. \ref{Fig: T2 dist}.

Monte Carlo simulations are also used to calculate the average gate infidelity of the pulses shown in Fig.~\ref{Fig: All Pulses}. Fluctuations in two-qubit parameters $(\Omega_{xx},\Omega_{yy})$ must now be captured along with single-qubit parameters $(\Omega_x^{(1)},\Omega_y^{(1)},\Omega_x^{(2)},\Omega_y^{(2)},a)$. Fluctuations in the two-qubit terms arise due to fluctuations of the interdot distance $L$, given by $\delta L=\delta r^{(1)}-\delta r^{(2)}$, where the $i^{th}$ dot's shift is given by
\begin{equation}
    \delta r^{(i)}= \frac{e}{m^* \omega_0^2} \delta F^{(i)},
    \label{Eq: Dot Shift}
\end{equation}
and $\delta F^{(i)}$ is the local electric field fluctuation at the $i^{th}$ dot. $L$ is less sensitive to motion orthogonal to the connecting axis between dots than parallel motion. Therefore, only parallel fluctuations in $L$ are calculated in the Monte Carlo simulation and $\delta F^{(i)}$/$\delta r^{(i)}$ can be considered scalars. Fluctuations in $(\Omega_{xx},\Omega_{yy},a)$ are different from the other parameters because they originate from $\delta L$. We calculate their fluctuations simply as $\delta \Omega_{xx}\approx\frac{d\Omega_{xx}}{dL} \delta L$, $\delta \Omega_{yy}\approx\frac{d\Omega_{yy}}{dL}\delta L$, and $\delta a \approx\frac{da}{dL}\delta L$, assuming $\delta L$ is small.

The Monte Carlo procedure for calculating pulse infidelities again begins by randomly sampling the configuration space of dipole locations. This produces perturbations $(\delta \Omega_x^{(1)},\delta \Omega_y^{(1)},\delta \Omega_x^{(2)},\delta \Omega_y^{(2)},\delta \Omega_{xx},\delta \Omega_{yy})$ through the first and second derivatives of the noise potential instance at both dots. We compute the unitary evolution due to each perturbed pulse and evaluate the infidelity. This process is repeated $N$ times, after which we average each pulse's infidelity. 

The result of a single Monte Carlo simulation with $N=1000$ predicts the pulse infidelities in Table \ref{Table: avg inf}. Generally, gates with $B\neq0$ have less infidelity because those pulses require less control parameters affected by charge noise.
\begin{table}[h!]
\centering
\begin{tabular}{c c c} 
 \hline
 Gate  & $f_c = 4.5$ GHz & $f_c = 0$ GHz\\ [0.5ex] 
 \hline
  bSWAP& $0.6$ & $1$ \\ 
  HI& $0.6$ & $6$ \\ 
  TI& $0.5$ & $3$ \\
  SI& $0.9$ & $3$ \\
  II& $0.4$ & $2$ \\
  \hline
\end{tabular}
\caption{Values of average infidelity $\braket{1-\mathcal{F}} \times 10^{5}$ for all eight GRAPE optimized pulses in the presence of single and two qubit noise. }
\label{Table: avg inf}
\end{table}

\section{III. \ Analytical Charge Noise Calculations}\label{Sup: Analytic Noise}
For a given dipole configuration, the standard deviation (SD) of voltage $\sigma_V$ and electric field $\sigma_F$ can be calculated exactly. Consider a point dipole whose center of mass coordinates are $(x_i,y_i,h)$, with a dipole vector parallel to the $z=h$ plane. The angle made by the dipole vector and the $x$ axis is $\phi_i$. The potential produced by the dipole at the point $(x,y,0)$ is given by
\begin{equation}\label{Eq: Dip Potential}
    V_{dip}(\vec{r},\vec{r}_i,\phi_i)=\frac{qd}{4 \pi \epsilon_r \epsilon_0} \frac{(x-xi) \cos(\phi_i)+(y-yi) \sin(\phi_i)}{\left((x-xi)^2+(y-yi)^2+h^2\right)^{3/2}}
\end{equation}
where $q$ is the charge of the dipole and $d$ is the dipole length. The entire noise potential, for $n$ dipoles, is given by
\begin{equation}\label{Eq: V noise}
    V(\vec{r},\vec{r}_1,...,\vec{r}_n)=\sum_{i=1}^{n}V_{dip}(\vec{r},\vec{r}_i,\phi_i)
\end{equation}
The SD of the potential can be calculated by the following equation
\begin{equation}\label{Eq: VRMS 1}
    \sigma_V^2=\braket{V^2}-\braket{V}^2=\frac{1}{(2\pi)^n A^n}\prod_{j=1}^{n} \iint_A dA_j \int_0^{2\pi} d\phi_j V(\vec{r},\vec{r}_1,...,\vec{r}_n)^2 - \left(\frac{1}{(2\pi)^n A^n}\prod_{i=j}^{n} \iint_A dA_j \int_0^{2\pi} d\phi_j V(\vec{r},\vec{r}_1,...,\vec{r}_n)\right)^2.
\end{equation}
Integrating first over the $\phi$ variables, $\braket{V^2}$ has inter-dipole cross terms that integrate to zero while $\braket{V}^2$ vanishes entirely. We are left with
\begin{equation}\label{Eq: VRMS 2}
    \begin{aligned}
        \sigma_V^2&=\frac{1}{(2\pi)^n A^n}\prod_{j=1}^{n} \iint_A dA_j \int_0^{2\pi} d\phi_j \sum_{i=1}^n V_{dip}(\vec{r},\vec{r}_i,\phi_i)^2 \\
        &=\frac{1}{2 A^n} \left(\frac{qd}{4 \pi \epsilon_r \epsilon_0}\right)^2\sum_{i=1}^n \prod_{j=1}^{n} \iint_A dA_j \frac{(x-x_i)^2+(y-y_i)^2}{\left((x-x_i)^2+(y-y_i)^2+h^2\right)^3}
    \end{aligned}
\end{equation}
Switching to polar coordinates and evaluating Eq.~\eqref{Eq: VRMS 2} at the origin, we get
\begin{equation}\label{Eq: VRMS 3}
    \begin{aligned}
        &\sigma_{V_0}^2= \left(\frac{qd}{4 \pi \epsilon_r \epsilon_0}\right)^2 \frac{n r_0^2}{4h^2(h^2+r_0^2)^2},
    \end{aligned}
\end{equation}
where $r_0$ is the radius of the area $A$ integrated over. Taking the derivative of Eq.~\eqref{Eq: Dip Potential} with respect to a single direction and following the same procedure outlined above, one can calculate the RMS fluctuations in the electric field at the origin ($F_0$). The result is

\begin{equation}\label{Eq: ERMS}
    \sigma_{F_0}^2=\left(\frac{qd}{4 \pi \epsilon_r \epsilon_0}\right)^2 \frac{n\left(8 h^6 + 8 h^4 r_0^2 + 12 h^2 r_0^4 + 3 r_0^6\right)}{16 h^4 (h^2+ r_0^2)^4}.
\end{equation}

It is now straightforward to calculate the dephasing of an EDSR qubit. The SD of the qubit's energy splitting is
\begin{equation}\label{Eq: EDSR energy RMS}
    \sigma_E = g \mu_B b_{long}\sigma_r = g \mu_B b_{long}\frac{e \sigma_{F_0}}{ m^* \omega_0^2},
\end{equation}
where $b_{long}$ is the longitudinal magnetic field gradient, $\mu_B$ is the Bohr magneton and $g$ is the electron g-factor. The last equality of Eq.~\eqref{Eq: EDSR energy RMS} comes from Eq.~\eqref{Eq: Dot Shift}. Using Eq.~\eqref{Eq: EDSR energy RMS} to calculate $T_2^*$ we arrive at
\begin{equation} \label{Eq: T2 EDSR}
    T_2^*=\frac{2\sqrt{2}}{g} \frac{m^* m_e \omega_0^2 }{e^2 b_{long}\sigma_{F_0}}.
\end{equation}
Given $b_{long} = .2 \text{ }\frac{m \text{T}}{\text{nm}}$ \cite{Yoneda2018} and $ \omega_0 = 2 \text{ meV}/\hbar$, along with all the noise model parameters used in the main text, $T_2^* \approx 20 \text{ }\mu$s.

\section{IV. \ Relaxation Due to Phonons}\label{Sup: Relaxation}
The relaxation time within the $p$-orbital subspace can be estimated using the framework of Ref.~\cite{TahanRelaxation2014}. Consider the phonon Hamiltonian,
\begin{equation}
    H_{ep}=\sum_{\lambda=1}^3\Sigma_\textbf{q}i \left[a_{\textbf{q}\lambda}^*\text{exp}[-i\textbf{q}\cdot\textbf{r}] + a_{\textbf{q}\lambda}\text{exp}[i\textbf{q}\cdot\textbf{r}]\right]\times\left[ \Xi_d\textbf{e}(\textbf{q},\lambda)\cdot \textbf{q}+\Xi_u e_z(\textbf{q},\lambda)q_z\right],
\end{equation}
where $\Xi_d$, $\Xi_u$ are matrix elements of the deformation potential, $\textbf{q}$ is the phonon wavevector with polarization $\lambda$, and $\textbf{e}(\textbf{q},\lambda)$ is the unit displacement vector due to the phonon. One can calculate the relaxation time due to the emission of a phonon given Fermi's Golden Rule with the matrix element $\bra{\psi_{j,k},n_{\textbf{q},\lambda}}H_{ep}\ket{\psi_{j',k'},n_{\textbf{q},\lambda}+1}$, where $\psi_{j,k}$ is the two-dimensional harmonic oscillator wavefunction with $j$ excitations along $x$ and $k$ excitations along $y$. From the appendix of Ref.~\cite{TahanRelaxation2014}, we get the relaxation rate

\begin{equation}\label{Eq: Gamma}
\begin{aligned}
    &\frac{1}{T_1}=\frac{(n_q+1)}{ 2(2\pi^2)v_\lambda^2\rho_{Si} \hbar}\sum_\lambda q_{\Delta \lambda}\int_0^{2\pi}\int_0^\pi \sin(\theta)d\theta d\phi\left\lvert\left[\Xi_d q_{\Delta l} + \Xi_u e_z(\textbf{q}_{\Delta \lambda},\lambda)q_z\right]f(\textbf{q}_{\Delta \lambda})\right\rvert^2
\end{aligned}
\end{equation}
where, generally, for a transition between two states $\psi_{n,m}$ and $\psi_{n',m'}$ due to the emission of a phonon with wavevector $\textbf{q}$,
\begin{equation}\label{Eq: f}
    f(\textbf{q})=\int d\textbf{r} \psi_{j,k}(\textbf{r}) \text{exp}(i \textbf{q} \cdot \textbf{r})\psi_{j',k'}(\textbf{r}).
\end{equation}
Now, consider relaxation within the $p$-orbital manifold. Eq.~\ref{Eq: f} becomes 
\begin{equation}\label{Eq: pxpy}
\begin{aligned}
    f(\textbf{q})&=\iint_{-\infty}^{\infty}dxdy\psi_{0,1}(x,y) \text{exp}[i\left(q_x x+ q_y y\right)]\psi_{1,0}(x,y)\int_{-\infty}^{\infty}dz\text{exp}[i q_z z]\left(\frac{2}{\pi z_0^2}\right)^{1/4}\text{exp}[-\frac{z^2}{z_0^2}]\\&=-\frac{\hbar \sqrt{r} q_x q_y}{2 m \omega_x}\text{exp}\left[-\frac{\hbar}{4 m \omega_x}\left(q_x^2 + r q_y^2\right)\right]\text{exp}[-\frac{1}{8}q_z^2 z_0],
\end{aligned}
\end{equation}
where $\omega_x$ is the confinement frequency along the $x$ direction and $r=\frac{\omega_x}{\omega_y}=\frac{E_x}{E_y}$ is the fixed ratio of confinement energies along the two dot axes. Converting $\textbf{q}$ to spherical coordinates and setting $q\rightarrow \frac{\Delta E}{\hbar v}$ with $\Delta E=E_x-E_y$, one can rewrite Eq.~\ref{Eq: pxpy} as
\begin{equation}\label{Eq:fintegrated}
    f(\textbf{q})=-\frac{ \sqrt{r} (1-1/r) \cos(\theta)^2 \sin(\phi)\cos(\phi) \Delta E }{2 m v^2}\text{exp}\left[-\frac{\hbar}{4 m \omega_x}\left(q_x^2 + r q_y^2\right)\right]\text{exp}[-\frac{1}{8}q_z^2 z_0] \propto \Delta E
\end{equation}
Plugging this into Eq.~\ref{Eq: Gamma}, we get a relaxation rate
\begin{equation}
    \frac{1}{T_1} \propto (\Delta E)^5.
\end{equation}
This scaling behavior tells us that the relaxation time increases as the transition energy decreases. This behavior is also seen in Ref.~\cite{TahanRelaxation2014}, but with a weaker dependence on $\Delta E$. Given a dot with $1-r \sim .01$ (1\% deformation), $E_x \sim 4 \text{ meV} \rightarrow l_x \sim 10 \text{ nm}$, and assuming all physical constant values taken in Ref.~\cite{TahanRelaxation2014}, Eq.~\eqref{Eq:fintegrated} gives $T_1 \sim 1 \text{ second}$.

\section{V. \ Pulse Optimization}\label{Pulse}
\begin{figure}[h]
    \centerline{\includegraphics[scale=.38]{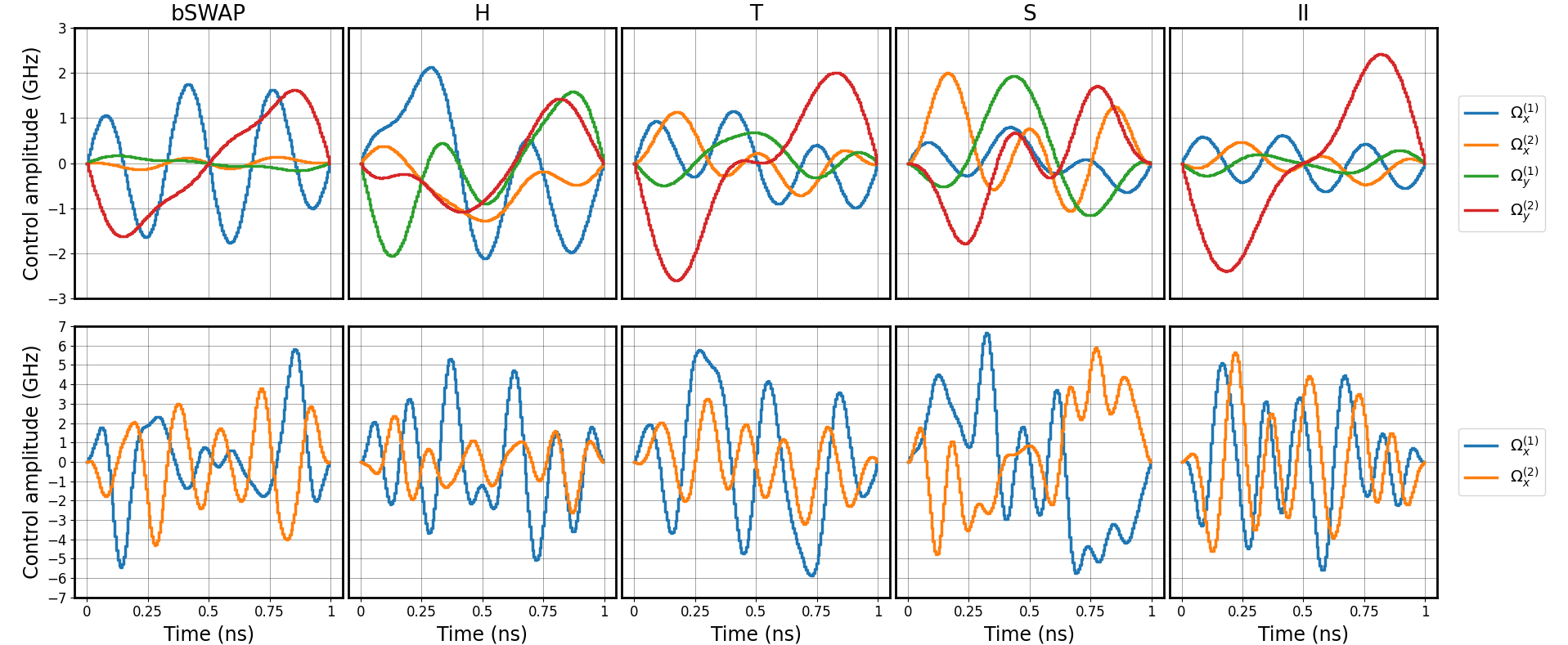}}
    \caption{GRAPE optimized pulse shapes. Upper half (lower half) corresponds to $\frac{\omega_c}{2\pi} = 0\text{ }(4.5)$ GHz and a maximum available bandwidth of $3 \text{ }(10)$ GHz. The legend next to each half indicates what control parameters were used for the respective pulses.}
    \label{Fig: All Pulses}
\end{figure}

Control pulses for elements of the universal gate set are created using QuTiP version 5.0.2 \cite{johansson2012qutip} and the Qutip quantum information processing package QuTiPqtrl. The collection of pulses found with QuTiP are shown in Fig.~\ref{Fig: All Pulses}.
The exact code used to generate these pulses can be found in this paper's companion repository \cite{GitHub}, along with updated modules that should replace their predecessors in the standard QuTip-qtrl package. The module edits enable 1) a Fermi function envelope to GRAPE generated pulses and 2) a high frequency contribution to the cost function. Explicit details about how to run the code can be found in the repository's README.

\putbib[ref_clean.bib]
\end{bibunit}

\bibliography{ref_clean}